\documentclass[sigconf,screen]{acmart}

\AtBeginDocument{%
  }

\title{Prompting in Practice: Investigating Software Practitioners’ Use of Generative AI Tools}

\author{Daniel Otten, Trevor Stalnaker, Nathan Wintersgill, Oscar Chaparro, Denys Poshyvanyk}
\email{{dsotten, twstalnaker, njwintersgill, oscarch, dposhyvanyk}@wm.edu}
\affiliation{%
 \institution{William \& Mary, Williamsburg, VA}
 \country{USA}
}

\date{July 2025}

\usepackage{ifthen}

\newboolean{showcomments}
\setboolean{showcomments}{true}

\newboolean{showboxes}
\setboolean{showboxes}{true}

\usepackage{verbatim}

\usepackage[utf8]{inputenc}
\usepackage{url}
\usepackage{colortbl}
\usepackage{balance}
\usepackage{xspace}
\usepackage{graphicx}
\usepackage{bold-extra}
\usepackage{paralist}
\usepackage{multirow}
\usepackage{listings}
\usepackage{boxedminipage}
\usepackage{soul}
\usepackage{enumitem}
\usepackage[normalem]{ulem}
\setlist{nolistsep,leftmargin=.5cm}
\usepackage{booktabs}
\usepackage{tabularx}
\usepackage{svg}
\usepackage{calc}
\usepackage{float}
\usepackage{multirow}
\usepackage[normalem]{ulem}
\useunder{\uline}{\ul}{}
\usepackage[most]{tcolorbox}
\usepackage{caption}
\usepackage{microtype} 
\usepackage{algorithmic}
\usepackage{textcomp}
\usepackage[dvipsnames]{xcolor}

\usepackage{wrapfig}
\usepackage{subcaption}
\usepackage{cleveref} 
\usepackage{adjustbox}
\usepackage{totcount}

\usepackage{tikz}
\usetikzlibrary{shapes.geometric, arrows.meta, positioning, decorations.pathmorphing}

\usepackage{breakurl}

\usepackage{hyperref}

\ifthenelse{\boolean{showcomments}}
{\newcommand{\nb}[2]{
		\fbox{\bfseries\sffamily\scriptsize#1}
		{\sf\small$\blacktriangleright$\textit{#2}$\blacktriangleleft$}
	}
	
}
{\newcommand{\nb}[2]{}
	
}

\newcommand{\ie}{\textit{i.e.},\xspace}
\newcommand{\eg}{\textit{e.g.},\xspace}
\newcommand{\etc}{\textit{etc.}\xspace}
\newcommand{\etal}{\textit{et al.}\xspace}

\newcommand{\strategy}[1]{\textit{#1}\xspace}

\newcounter{findingcounter}
\newtotcounter{totalfindings}
\ifthenelse{\boolean{showboxes}}
{
    \newcommand{\finding}[1]{%
      \refstepcounter{findingcounter}
      \begin{tcolorbox}[boxsep=1pt,left=2pt,right=2pt,top=1pt,bottom=1pt]%
      \small
      \textbf{Finding \arabic{findingcounter}:} #1
      \end{tcolorbox}%
      \addtocounter{totalfindings}{1}
    }
    
  }
{
    \newcommand{\finding}[1]{}
}

\begin{abstract}

The use of generative AI (GenAI) tools has fundamentally transformed software development. Central to this shift is prompt engineering, the practice of crafting textual prompts to guide GenAI tools in generating useful content. Although prompt engineering has emerged as a critical skill, prior research has focused primarily on cataloging of prompting techniques, with limited attention to how software practitioners employ GenAI within real-world development workflows. To address this gap, this study presents a systematic investigation of  practitioners' integration of GenAI tools into software development, drawing on a rigorous survey that examines prompting strategies, conversation patterns, and reliability assessments across core software development tasks.
We surveyed 72 software practitioners who actively use GenAI to characterize AI usage patterns throughout the development process. By combining qualitative and quantitative analyses of the survey responses, we identified \total{totalfindings} key findings that describe how prompting is performed in practice. Our study shows that while code generation is nearly universal, proficiency strongly correlates with the use of GenAI for more nuanced tasks such as debugging and code review. Practitioners also tend to favor iterative multi-turn conversations to single-shot prompting. Documentation tasks are perceived as most reliable, while complex code generation and debugging remain major challenges. Our findings provide an empirical view of practitioner practices, ranging from basic code generation to deeper integration of GenAI into development workflows, enabling us to offer recommendations for improving both GenAI tools and the ways practitioners interact with them.

\end{abstract}

\begin{document}

\maketitle

\section{Introduction}
\label{sec:intro}

Generative Artificial Intelligence (GenAI) is rapidly transforming the way programmers and developers create and evolve software. With the help of GenAI tools such as GitHub Copilot~\cite{copilot}, Windsurf~\cite{codeium}, Claude Code~\cite{claude-code}, and ChatGPT~\cite{chatgpt}, software practitioners can now easily and quickly generate new code, review and modify existing codebases, identify and correct defects, and produce code documentation. Central to these capabilities is prompting: the practice of crafting natural language instructions to guide GenAI systems toward producing useful and accurate output. Prompting has emerged as a novel engineering practice, fundamentally different from traditional software development due to its reliance on large language models (LLMs), which are inherently probabilistic and non-deterministic, but highly capable~\cite{atil2025nondeterminismdeterministicllmsettings}.

Previous research has primarily focused on cataloging prompting techniques~\cite{white2023promptpatterncatalogenhance, schulhoff2025promptreportsystematicsurvey, sahoo2025systematicsurveypromptengineering}, 
which offers limited information on how practitioners actually use prompting in practice: how they craft prompts for different types of software engineering (SE) tasks, how they adapt to unreliable or suboptimal output, and how GenAI is reshaping standard development activities.  
Previous studies have examined narrow use cases (\eg refactoring~\cite{alomar2024refactor}), specific populations (\eg students~\cite{alpizar2025student, choudhuri2025insights}) and broader socio-organizational adoption factors~\cite{li2024ai, 10.1145/3652154}. However, there remains a lack of systematic investigation into real-world prompting practices for a variety of SE tasks, which is essential to  evaluate current practices, inform practitioners and educational guidance, and drive the design of more effective GenAI technology.
\looseness=-1

To address this gap, we conducted a survey of 72 software practitioners, exploring how they use GenAI tools and engage in prompting across six core SE tasks: code generation, documentation, debugging, testing, refactoring, and review. (We use \textit{software practitioners} to denote professionals with experience in programming and software development, including developers, architects, technical leads, researchers, and other roles.) Our study investigates multiple dimensions of GenAI usage, including the prompting and conversational strategies that practitioners employ, how they assess and respond to the quality of GenAI output, and how prompting practices are integrated into their daily workflows. Together, these dimensions provide a holistic view of how prompting functions in modern software development and extend the current understanding of prompt engineering in practice.

Our study identified \total{totalfindings} key findings of prompt engineering for software development. In particular,  we find that while nearly all practitioners who use GenAI apply it to code generation, higher perceived proficiency strongly correlates with engagement in more types of tasks, such as code refactoring and testing. Practitioners rely primarily on iterative conversations, with daily usage critical to achieving significant productivity improvements. Our reliability assessment reveals that practitioners trust GenAI more for documentation and least for complex code generation/modification tasks. These findings have important implications for practitioners seeking to improve productivity through effective prompting strategies, for tool builders designing more reliable and intuitive GenAI, and for educators developing curricula around the evolving skill set needed for effective GenAI collaboration.

This paper makes the following contributions: (1) a comprehensive characterization of how GenAI tools are used in various SE tasks; (2) empirical insights into practitioners’ prompting strategies, reliability perceptions, and common failure types; and (3) actionable implications for improving human-AI interaction in SE.

\section{Study Methodology}
\label{sec:methodology}

The \textit{goal} of our study is to investigate how software practitioners use prompt engineering and GenAI tools for development tasks and what challenges they face during this use. The \textit{context} consists of software practitioners with experience using GenAI tools for SE.

Our study addresses the following research questions (RQs): 
\begin{enumerate}[label=\textbf{\labelitemi \space RQ$_\arabic*$:}, ref=\textbf{RQ$_\arabic*$}, wide, labelindent=5pt, leftmargin=5pt]\setlength{\itemsep}{0.2em}
    \item \label{rq:1} How do software practitioners integrate GenAI tools into their development workflows?
    \item \label{rq:2} What prompting techniques and conversation strategies do practitioners use when interacting with GenAI tools? 
    \item \label{rq:3} How reliable do practitioners perceive GenAI to be for various SE tasks and what are common issues they encounter?
\end{enumerate}

\begin{figure}[t]
    \centering
    \includegraphics[width=\linewidth]{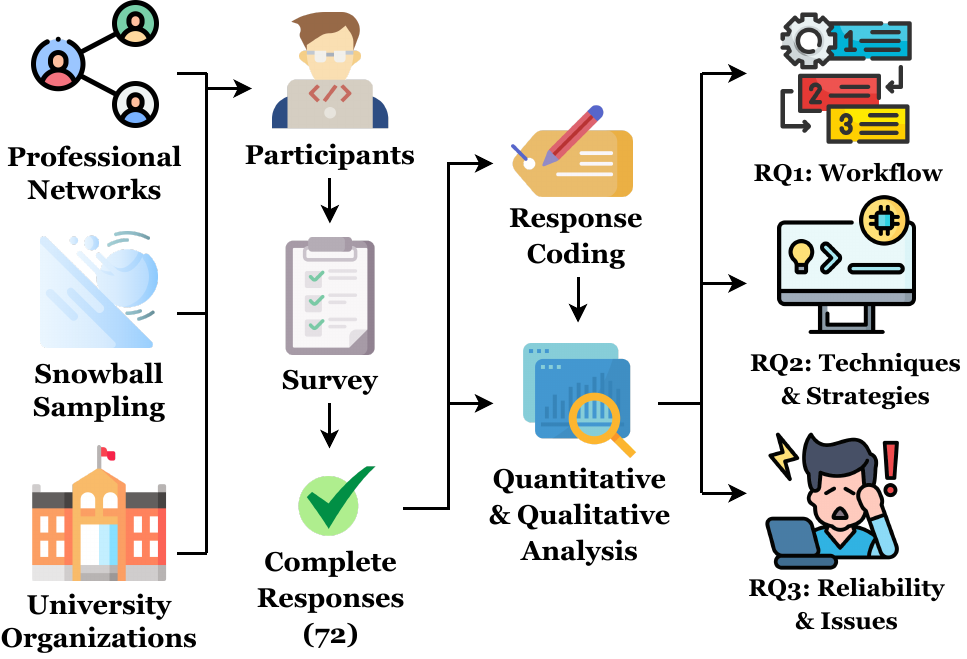}
    \caption{Study Methodology (image credits at \cite{anonymous_repo})}
    \label{fig:methods}
\end{figure}

Next, we detail our methodology (\Cref{fig:methods}), which aligns with our previous studies~\cite{stalnaker2024boms,stalnaker2024developer,wintersgill2024law,zappin2025challenges} and was approved by our institution's ethics review board, including questionnaire, participant identification procedure, and data collection/distribution.

\subsection{Survey Design}
\label{sec:survey_design}

We designed our survey questionnaire considering the existing literature on prompt engineering and the use of GenAI tools~\cite{white2023promptpatterncatalogenhance,schulhoff2025promptreportsystematicsurvey}%
, following general guidelines~\cite{survey} and SE-specific guidelines for survey design~\cite{DBLP:journals/sigsoft/PfleegerK01,DBLP:journals/sigsoft/KitchenhamP02,DBLP:journals/sigsoft/KitchenhamP02a,DBLP:journals/sigsoft/KitchenhamP02b,DBLP:journals/sigsoft/KitchenhamP02c,DBLP:journals/sigsoft/KitchenhamP03,linaaker2015guidelines,molleri2020empirically}.
Through the survey, our goal was to build an understanding of GenAI integration across six core SE tasks: %
code generation, refactoring, testing,  debugging, documentation, and code review. 
We selected these tasks because they span diverse development activities, are commonly supported by GenAI tools, and vary in complexity and risk, allowing a broad examination of prompting practices. This selection also balances task coverage with survey length, as we included task-specific questions.
We implemented branching logic in the survey, so that participants were only shown %
questions relevant to their experiences (\eg SE tasks for which they have used GenAI).  
To ensure data quality, %
we also included an initial question to screen out respondents with no experience using prompt-based GenAI tools. %
We carefully formulated the survey questions, ensuring that they were written clearly to avoid confusing or biasing the respondents. We also conducted a pilot study with graduate students from our research lab who regularly use GenAI tools. Based on their feedback, we further refined our questionnaire, improving questions that were unclear or confusing.
\looseness=-1

The survey was organized into five sections. Sec. 1 collected demographic information. Sec. 2 asked questions related to the GenAI experience of the participants, the GenAI tools they regularly use, and their GenAI development workflows.  Sec. 3 and 4 investigated the use of prompting techniques and conversation strategies. %
The final section explored the perceived reliability of GenAI tools and the issues that practitioners face. The questions  were presented in a variety of formats, including multiple-selection, Likert scales, and open text fields. The full text and logic of our survey can be found in our replication package~\cite{anonymous_repo}.

\subsection{Participant Recruitment}
\label{sec:recruitment}

Our recruitment strategy targeted software professionals %
with diverse backgrounds, including developers, architects, technical leads, researchers, and other roles with programming and software development experience. %
For survey distribution, we mainly relied on our professional networks and snowball sampling (\eg participants sharing the survey within their networks) by directly contacting potential participants or posting participation calls on social media platforms (\eg LinkedIn and X). To further broaden our sample and include professionals in earlier career stages, we distributed the survey to relevant organizations at our university, including the ACM chapter. %
Participants were not compensated monetarily or otherwise. 

\subsection{Data Collection and Analysis}
\label{sec:analysis}

Survey responses were collected using Qualtrics~\cite{qualtrics} between 27 May and 7 July 2025. To analyze the data, we used a mixed method approach, combining quantitative and qualitative techniques.

For multiple choice and Likert scale questions, we calculated data distributions and descriptive statistics 
to identify trends and patterns. Where appropriate, to facilitate comparison, we assigned ordinal Likert scale responses (\eg "Often" to "Never") to integer values (\eg 1 to 4). %
This allowed us to compute mean scores to represent concepts such as the popularity of a prompting strategy or the frequency of an issue using GenAI tools.  
\looseness=-1

\begin{table*}[t]
\centering
\resizebox{\textwidth}{!}{
 \begin{tabular}{|l|c|c|c|c|c|c|c|} 
 \hline
 \textbf{GenAI Proficiency} & \textbf{Code Gen.} & \textbf{Refactoring} & \textbf{Testing} & \textbf{Debugging} & \textbf{Documentation} & \textbf{Code Review} & \textbf{Avg. \# of tasks} \\ \hline
 {Minimum Proficiency} & 1/1 (100\%) & 0/1 (0\%) & 0/1 (0\%) & 0/1 (0\%) & 0/1 (0\%) & 0/1 (0\%) & 1.00 \\ \hline
 {Somewhat Proficient}  & 11/14 (78.5\%) & 3/14 (21.4\%) & 6/14 (42.9\%) & 4/14 (30.8\%) & 4/14 (30.8\%)    &  3/14 (23.1\%)  & 2.21        \\ \hline
 {Proficient}  & 32/32 (100\%) & 11/32 (34.4\%) & 9/32 (28.1\%) &  16/32 (50\%) & 14/32 (43.8\%)    & 8/32 (25.0\%)  & 2.81    \\ \hline 
 {Very Proficient}     & 15/18 (83.3\%) & 6/18 (33.3\%)  & 8/18 (41.2\%) & 8/18 (41.2\%) & 11/18 (58.8\%)  &  8/18 (41.2\%)  & 3.11     \\ \hline
 {Maximally Proficient}   & 7/7 (100\%) & 3/7 (42.9\%) & 5/7 (71.4\%) & 4/7 (57.1\%) & 5/7 (71.4\%)  & 4/7 (57.1\%)  & 4.00       \\ \hline
 {\textit{Total}} & 66/72 (91.7\%) & 23/72 (31.9\%) & 28/72 (38.9\%) & 32/72 (44.4\%) & 34/72 (47.2\%) & 23/72 (31.9\%) & 2.86   \\ \hline
 \end{tabular}
}
\caption{Proportion of respondents in each self-reported proficiency level who indicated GenAI use for each SE task}
\label{table:skill-vs-tasks}
\end{table*}

The responses to open-ended responses were analyzed using a qualitative coding approach~\cite{spencer2009card}. Three annotators (\ie authors) independently analyzed such responses 
and performed \textit{open-coding} by assigning one or more \textit{codes} to each response using a shared codebook (\ie a spreadsheet).  After the initial coding phase, the annotators met to collaboratively review, discuss, and reconcile the coding decisions.  Through this process, the annotators resolved any discrepancies, which included differing interpretations of potentially ambiguous responses, applying the same codes in different contexts, and using broader codes rather than more specific ones. 
Ultimately, this process resulted in a final set of codes, definitions, and labels for each response, which are found in our replication package~\cite{anonymous_repo}. Since codes were developed iteratively through an inductive process without an initial codebook and multiple codes could be assigned to each response, we did not base our analysis on inter-annotator agreements. We followed best open-coding practices \cite{spencer2009card} and used annotator discussions to ensure the reliability of the results.
\looseness=-1

\subsection{Participant Demographics}
\label{sec:demographics}

Our survey attracted  
91 respondents who completed the survey, of whom 72 (79\%) indicated having experience with GenAI tools and prompting. 
We answer our RQs based on these 72 responses.

The 72 participants represent a diverse, global sample of software professionals, with reported development experience of 1-40 years (mean = 10.2, median = 10). They held a variety of roles, including researcher (64\%), programmer (60\%), technical lead (28\%), and software architect (28\%). Respondents could select multiple roles, and overlaps were common: for example, 31 of 46 participants identified as researchers (the most frequently selected role) also reported other roles, notably programmer, architect, technical lead, and/or tester. For comparative analysis, we grouped professional experience into four categories: 24 \textit{juniors} (1–5 years), 22 \textit{intermediates} (6–10 years), 19 \textit{seniors} (11–20 years), and 7 \textit{veterans} (21+ years).
\looseness=-1

Participants reported having developed software across 43 distinct domains, with the most common being Healthcare (22\%), Research (17.6\%), Banking (15.4\%), Software Development (12.4\%), and Communication (10.1\%). When asked to choose from a predefined list of eight system types, they most frequently indicated experience with web applications (63.9\%), desktop applications (33.3\%), libraries/frameworks (31.2\%), and AI-intensive systems (30.6\%).
Respondents were located in 16 countries, with the majority residing in the United States (54\%), followed by Canada (11\%) and the United Kingdom (8\%). The most commonly used programming languages were Python (83\%), Java (61\%), and JavaScript (39\%).

\section{Results}
\label{sec:results}

This section presents the results for each RQ, while \cref{sec:disc} presents a discussion and implications of these results.

\subsection{\ref{rq:1}: GenAI Usage Patterns}

We first present the results practitioners' GenAI usage patterns.

\subsubsection{GenAI Use for SE Tasks and Proficiency} 

\Cref{table:skill-vs-tasks} shows that the vast majority of respondents (91.7\%) regularly use GenAI for code generation. Although this might be expected, we found that debugging (47.2\%) and code documentation (44.4\%) emerged as the next most common tasks. The remaining tasks, testing (38.9\%, refactoring (31.9\%), and code review (31.9\%) were less frequently reported. In general, 89\% of the respondents reported using GenAI for four or fewer tasks, with two (30.6\%) or three (23.6\%) tasks being most common, suggesting a relatively focused use of GenAI.

When analyzing the 5-Likert point responses of self-reported \textit{proficiency} in using GenAI, we found that the majority of participants (57 of 72) considered themselves proficient or very/maximally proficient (see \Cref{table:skill-vs-tasks}). (Since a single veteran practitioner rated themselves as having ``minimum proficiency'', we exclude this data point from future comparisons across proficiency levels.)

We also observed that proficiency correlates with the number of SE tasks for which participants use GenAI (Spearman's $\rho$ = 0.325). Among those who reported using GenAI for four or more SE tasks, most (18 of 22, 82\%) identified as very or maximally proficient. By contrast, single-task users reported being somewhat proficient or proficient (7 of 11, 64\%). These results suggest that task breadth and GenAI proficiency may mutually reinforce one another. 

When examining proficiency by software development experience, we found that junior and intermediate practitioners considered themselves more proficient in using GenAI that seniors and veterans. Specifically, 22 of 24 juniors  and 19 of 22 intermediates reported being proficient or very/maximally proficient, compared to only 12 of 19 seniors and 4 of 7 veterans.

\finding{A moderate correlation between perceived GenAI \textit{proficiency} and GenAI \textit{use} across six SE tasks suggests that that greater proficiency is associated with broader usage, and vice versa. Junior and intermediate practitioners tend to rate themselves as more proficient in using GenAI compared to senior and veteran practitioners.}

\looseness=-1

\subsubsection{GenAI Usage Frequency and Tool Adoption}
\label{sec:productivity}

Over 70\% of respondents reported engaging with GenAI tools once or multiple times per day. Daily usage increases with proficiency: 35.7\% of somewhat proficient users (5) reported daily GenAI use, as did 78.1\% of proficient users (25), 83.3\% of very proficient users (15), and 100\% of maximally proficient users (7).%

\begin{figure}[t]
    \centering
    \includegraphics[width=\linewidth]{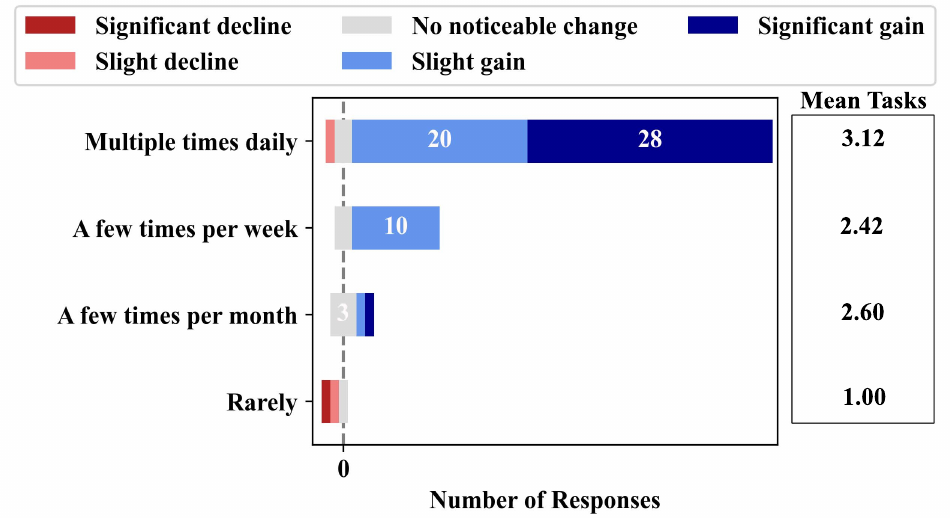}
    \caption{Productivity and task breadth by usage frequency}
    \label{fig:freq-productivity}
\end{figure}

Respondents who reported using GenAI more frequently tended to also report using GenAI for more SE task and experiencing productivity gains (\Cref{fig:freq-productivity}). Daily users reported using GenAI for 3.12 tasks on average, 
compared to 2.42 tasks for weekly user (an increase of 29\%). Reported usage frequency was moderately correlated with the breadth of GenAI use across tasks (Spearman’s $\rho$ = 0.348). %
Compared to non-daily users, daily GenAI users were also substantially more likely to state that GenAI significantly increased their productivity, with only one respondent who used GenAI tools non-daily reported such substantial gains.  Taken together, these results suggest that sporadic GenAI use is insufficient to realize meaningful productivity gains and that users who perceive higher productivity gains are, in turn, more likely to use such tools often.

\finding{Most respondents (69.3\%) reported using GenAI for SE tasks once or multiple times per day. In contrast, only 35.7\% of those who considered themselves somewhat proficient reported daily GenAI usage, suggesting a link between consistent use and perceived proficiency. Respondents who used GenAI more frequently also reported applying it across a larger number of SE tasks.}

Analyzing tool adoption from a given predefined list of options (with an open-ended "Other(s)" option), OpenAI's ChatGPT led reported tool adoption (83.1\%), 
followed by GitHub Copilot (43.7\%), Google's Gemini (35.2\%), and Anthropic's Claude (32.4\%). Open-weight 
models such as Deepseek-R1 (12.7\%) and the Llama family of models (7\%) showed more modest adoption.  The use of AI-integrated development environments (besides GitHub Copilot) was reported by 19.7\% of the participants. Examples include Cursor (6), Windsurf (1), and JetBrains AI Assistant (1).

\finding{The vast majority of practitioners (83.1\%) reported using ChatGPT for SE tasks, which was nearly 2× higher than the next most popular tool (GitHub Copilot with 43.7\%), suggesting that it has established itself as the leading platform for AI-assisted software development. 
}

When interacting with GenAI tools for SE tasks, an equivalent number of participants reported using tools with a web interface (46, 63.9\%) and IDE-integrated solutions (45, 62.5\%). Other GenAI integrations such as GenAI APIs (14, 19.4\%) and command line tools showed lower adoption (13, 18.3\%). Despite having used GenAI for development tasks, five respondents (6.9\%) indicated they had not been incorporated it into their workflows. 

A majority of practitioners (39, 54.2\%) reported integrating GenAI into their workflows in two or more ways, while 28 (38.9\%) reported a single integration, most commonly IDE-based (13) or web-based (12). Among participants using two integration types, 66.7\% (20) reported using both web-based and IDE-integration. These findings suggest that while command-line and API-based GenAI integrations are available, they are rarely used in isolation by practitioners. \Cref{table:Integration Sophistication Updated} summarizes the \# of participants across GenAI tool integrations. (No "Other" integrations were reported by participants.)

\begin{table}[t]
\centering
\small
\begin{tabular}{|c|c|c|}
\hline
\textbf{GenAI Integration} & \textbf{\# of Users} & \textbf{Percentage} \\ \hline
No integration &  5 & 6.9\% \\ \hline
Single integration & 28 & 38.9\% \\ \hline
Dual integration & 30 & 41.7\% \\ \hline
Triple integration&  6 & 8.3\% \\ \hline
Four integrations &  3 & 4.2\% \\ \hline
\end{tabular}
\caption{Distribution of participants by number of GenAI workflow integrations (Web, IDE, Command line, and API)}
\label{table:Integration Sophistication Updated}
\end{table}

Respondents that reported daily GenAI usage also demonstrated a higher number of workflow integration types when compared with weekly users: 60.9\% (31/51) reported used two or more integration methods, compared to 25\% (3/12) for weekly users. 

\finding{Practitioners reported no preference between incorporating web- or IDE-based GenAI tools into their development workflows, but did overwhelmingly employ one or both over command-line and API-based tools. Those who reported using GenAI tools multiple times a day were also more likely to incorporate GenAI tools into their workflows in multiple ways (Web-based, IDE-integration, \etc). This suggests that frequent GenAI users may develop more tailored or complex workflows.
\looseness=-1
}

\subsubsection{Changes in Development Approaches with GenAI}

When asked if GenAI tools had changed their approach to software development, the majority of respondents (55/71, 77.5\%) indicated that they had experienced at least some change, while 7.0\% were unsure and 15.5\% reported no change. We asked respondents who reported experiencing changes to their development approach to elaborate on those changes. %
Nine participants reported using GenAI tools for tasks that previously required internet searches (\eg finding library documentation or searching Stack Overflow for code examples), and four reported using GenAI to assist in information retrieval tasks within their code base. Six indicated that GenAI can provide a starting point for projects, such as helping with ideation (3), initial project design (5), or writing boilerplate code (6). Five respondents noted that GenAI allowed them to shift development focus from repetitive coding tasks to ``higher-level'' tasks such as architecture design, and four respondents claimed that GenAI tools allowed them to write more documentation. Still others indicated that GenAI helps produce cleaner, more organized code (3), reduces the focus and memory required to complete coding tasks (2), allows for development with unfamiliar programming languages (2), and can assist in project configuration (2).

GenAI tool adoption has also created new processes in practitioner workflows. For example, six respondents spend more time reviewing and verifying AI-produced code. Four focus more on writing good prompts, while three have incorporated interactive prompting processes into their workflows.

\finding{Practitioners reported that GenAI has transformed their development workflows in diverse ways, ranging from replacing traditional search practices to supporting project design, reducing repetitive effort, and broadening programming capabilities, ultimately enabling a shift toward higher-level, more strategic software engineering tasks. In addition, participants noted that GenAI introduced new practices in their development workflow, such as validating AI-generated outputs and employing various prompt engineering strategies.}

\subsection{\ref{rq:2}: Prompting and Conversation Strategies}

We examine the specific prompting techniques and conversation strategies that software practitioners use when using GenAI tools. %

\subsubsection{Prompting Technique Familiarity}
\label{sec:technique_familiarity}

Participants were asked about three prompting techniques identified by White \etal (\strategy{Meta Language Creation}, \strategy{Output Automator}, and \strategy{Persona})~\cite{white2023promptpatterncatalogenhance} in addition to \strategy{Meta-Prompting} and \strategy{Few-Shot Learning} \cite{schulhoff2025promptreportsystematicsurvey}. We included two additional techniques, \strategy{Output Style} and \strategy{Condition Check}, which we did not find adequately represented in prior work. %
Technique definitions were provided to participants and can be found in \Cref{table:technique-desc}. 
\looseness=-1

\begin{table}[t]
\centering
\footnotesize %
 \adjustbox{width=\columnwidth,center}{%
\setlength{\tabcolsep}{2pt}
\begin{tabularx}{\columnwidth}{|l|X|}
\hline
\textbf{Technique} & \textbf{Description}                                         \\ \hline
Condition Check              & Produce certain output when specific condition(s) are met  \\ \hline
Few-Shot Learning            & Providing examples to guide generation                                \\ \hline
Meta Lang. Creation       & Adjusts the semantics of words, phrases, or symbols           \\ \hline
Meta-Prompting               & Having AI suggest better prompts for specific tasks                   \\ \hline
Output Automator             & Generating scripts to implement solutions                             \\ \hline
Output Style                 & Make the AI's output follow a particular format or style              \\ \hline
Persona                      & Ask AI to complete a task while acting as a certain role \\ \hline
\end{tabularx}%
}
\caption{Prompting technique descriptions}%
\label{table:technique-desc}
\end{table}

\begin{figure}
    \centering
    \includegraphics[width=\linewidth]{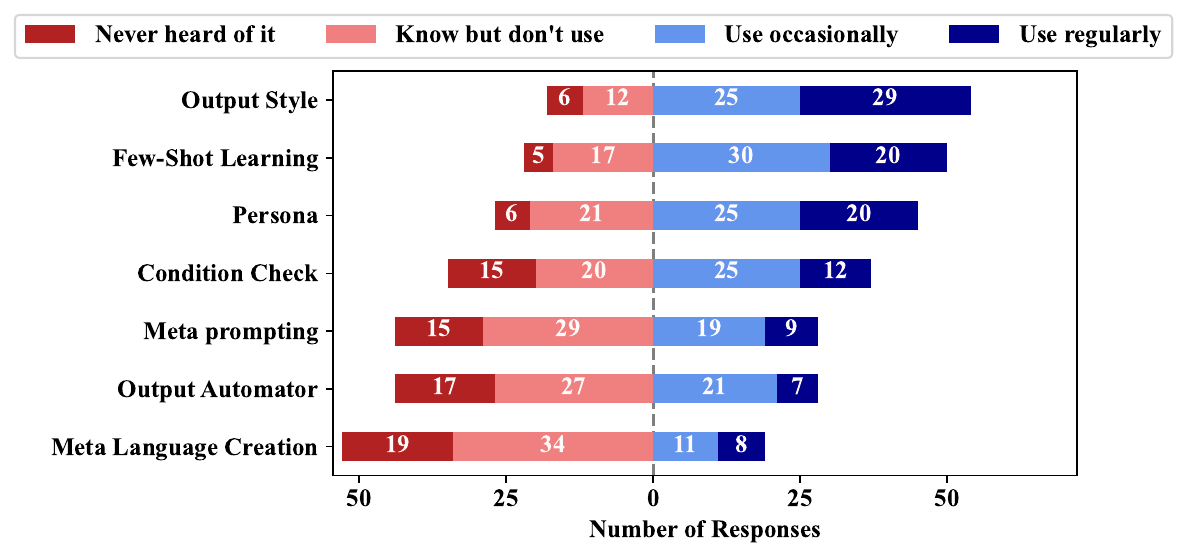}
    \caption{Prompting technique familiarity distribution}
    \label{fig:technique-familiarity}
\end{figure}

Our analysis of prompting technique familiarity reveals a clear hierarchy of adoption, as shown in \Cref{fig:technique-familiarity}. We consider participants were \textit{familiar} with a technique if they \textit{knew} or \textit{used} the technique (occasionally or regularly).
Respondents reported overwhelming familiarity with the aforementioned prompting strategies. 
Nearly all respondents were familiar with the \strategy{Few-Shot Learning} (93\%), \strategy{Output Style} (92\%), and \strategy{Persona} (92\%) techniques. The \strategy{Meta Language Creation} technique were the least familiar to participants (74\%). Despite the high familiarity, reported respondent usages (either occasionally or regularly) were markedly lower.  The \strategy{Output Style} (82\%), \strategy{Few-Shot Learning} (75\%), \strategy{Persona} (68\%), and \strategy{Condition Check} (65\%) strategies had the highest reported usage rates among respondents who were familiar with the respective strategies. Roughly half of respondents familiar with the \strategy{Output Automator} (51\%) and \strategy{Meta-Prompting} (49\%) strategies reported using them.  The \strategy{Meta Language Creation} strategy had the lowest adoption rate (36\%).
\looseness=-1

The \strategy{Output Style} (54\%) and \strategy{Persona} (44\%) strategies were  most likely to be used regularly, among participants who used them. Despite being  least popular overall, 42\% of respondents who used \strategy{Meta Language Creation}, use it regularly. Respondents most often reported using the \strategy{Output Automator} (75\%), \strategy{Condition Check} (68\%), and \strategy{Meta-Prompting} (68\%) strategies only occasionally.

\finding{Respondents were highly familiar with all seven prompting strategies, with 74\%–94\% reporting awareness. However, actual usage lagged behind, with only 36\%–82\% of those aware indicating that they used the techniques.}

\begin{table}[]
\footnotesize
\resizebox{\columnwidth}{!}{%
\begin{tabular}{|l|c|c|c|c|}
\hline
\textbf{Proficiency} & \textbf{
Familiar \%} & \textbf{Use \%} & \textbf{Occ. Use \%} & \textbf{Reg. Use \%} \\ \hline
Somewhat    & 74.5        & 52.1       & 89.5             & 10.5          \\ \hline
Proficient  & 83.9        & 60.1       & 62.8             & 37.2          \\ \hline
Very        & 89.7        & 69.0       & 51.3             & 48.7          \\ \hline
Maximally   & 95.9        & 68.1       & 34.4             & 65.6          \\ \hline
\textbf{Experience}             & \textbf{Familiar \%} & \textbf{Use \%} & \textbf{Occ. Use \%} & \textbf{Reg. Use \%} \\ \hline
Junior       & 85.1    & 58.0       & 67.5       & 32.5       \\ \hline
Intermediate & 87.7    & 63.7        & 45.4       & 54.7       \\ \hline
Senior       & 86.5    & 62.6       & 66.7       & 33.3       \\ \hline
Veteran      & 57.1    & 71.4       & 65.0          & 35.0          \\ \hline
\end{tabular}
}
\caption{Prompting technique familiarity and use by GenAI proficiency and SE experience categories}
\label{table:skill_vs_techniques}
\end{table}

None of respondents' self-reported GenAI proficiency, years of software development experience, or GenAI proficiency and experience impacted relative familiarity ratings of prompting techniques (\eg across proficiency groups, \strategy{Output Style} always had a higher familiarity than \strategy{Meta Language Creation}).

\Cref{table:skill_vs_techniques} reports the proportion of respondents who are familiar with, and use, the seven prompting techniques across GenAI proficiency and SE experience categories. Technique usage rates are further reported by \textit{occasional} and \textit{regular} use. The table reports total percentages by aggregating response counts across the techniques. The results show a clear pattern: both technique familiarity and usage (including occasional and regular use) consistently increase with self-reported proficiency. In other words, participants who considered themselves more proficient were also more likely to use prompting techniques more frequently. 

In contrast, analysis by practitioner experience reveals a different trend. With the exception of veterans, all experience groups show similar familiarity and usage rates. Veteran are the least familiar with prompting techniques, but among those who are familiar, they are the most likely to use them. Analyzing occasional versus regular technique usage, practitioners across all experience levels show consistent rates, except for intermediate practitioners, who reported over 60\% increase in regular usage compared with their peers. This suggests that intermediate practitioners may be at a career stage where they are both open to adapting their workflows and sufficiently experienced to integrate GenAI effectively.

\finding{Familiarity and usage of prompting techniques increase with GenAI proficiency rather than years of developement experience. Veteran practitioners are the least familiar with prompting techniques, yet the most likely to use the ones they know. While most experience groups use prompting at similar rates, intermediate practitioners stand out as especially active adopters, suggesting they are mid-career professionals that are uniquely positioned to integrate GenAI into their workflows.}

\subsubsection{Prompt Information Inclusion}

We describe the prompt information practitioners use for each of the six SE tasks we studied.
\looseness=-1

When prompting a GenAI tool for \textit{code generation}, 76\% of respondents (66) indicated that they often include example inputs and expected outputs.  Respondents also indicated including instructions for which libraries the model should use (53\%), a description of the existing codebase~(45\%), implementation alternatives to consider or avoid (41\%), error handling expectations (39\%), coding style guidelines (32\%), deployment constraints (32\%), and performance/optimization requirements (27\%).

When using GenAI tools for \textit{refactoring} tasks, respondents (23) most often include the original code with explanatory comments (78\%), a description of the existing issues or refactoring goals (65\%), and examples of similar code that they find to be well structured (61\%). Others opt to include architectural constraints and style guidelines (48\%), a specific refactoring pattern to apply (48\%), or code that should remain unchanged during the refactoring (43\%). The least common information supplied in respondent prompts include unit tests (22\%) and performance requirements (17\%).

Of the 28 respondents who use GenAI for \textit{testing} tasks, 82\% prompt the model with the code to be tested along with specifications. Respondents also include example test cases (46\%), testing preferences (43\%), information on edge cases or expected behaviors (39\%), existing test suite examples (39\%), and mocking instructions (36\%) in their prompts. The least  common to be included are test coverage goals (25\%), environmental requirements (25\%), and performance or resource constraints (14\%).

When using GenAI for \textit{debugging} tasks, nearly all respondents (94\%) provide the model with full error messages or stack traces. Similarly, 81\% often include related logs or console output.  Less common context given to the model includes environmental details (50\%), previously attempted solutions (44\%), (bug) reproduction steps (41\%), system architecture information (31\%), and visual evidence of issue manifestation (\eg screenshots) (25\%). Only two respondents (6\%) included version control history.

For \textit{documentation} tasks, respondents (34) reported that they often provide the model with the code snippet along with a functionality explanation (71\%). Some also provide examples highlighting the desired documentation style (53\%), usage examples/scenarios (41\%), context about the target audience (38\%), existing documentation in need of updates (38\%), format requirements (35\%), guidance on domain-specific terminology (26\%), or information on project standards and conventions (15\%).

Lastly, when using GenAI for \textit{code review}, respondents (23) predominantly provide the model with the code to be reviewed with some additional context (78\%). Beyond this, they include project coding standards (43\%), a description of the intended feature or purpose (35\%), or performance expectations (30\%), security/compliance requirements (22\%), warnings about common pitfalls in the domain or codebase (22\%), the code of related components (17\%), or feedback from previous reviews (17\%).

\subsubsection{Conversation Strategy Adoption}
\label{sec:conv_strats}

We examined how practitioners structure their conversations with GenAI tools. %
Participants rated the use frequency of nine conversation strategies we defined (seen in \Cref{table:conv-strategy-desc}) %
on a four-point Likert scale from ``Never'' to ``Often.'' %
This analysis reveals which strategies survey respondents prefer.

\begin{table}[]
\resizebox{\columnwidth}{!}{%
\begin{tabular}{|l|l|}
\hline
\textbf{Conversation Strategy} & \textbf{Description}                                        \\ \hline
Comparative analysis           & Having the AI generate multiple solutions for comparison    \\ \hline
Context building               & Adding more context as the conversation develops            \\ \hline
Exploratory dialogue           & AI explores multiple solutions before implementing one      \\ \hline
Feedback loop                  & Providing feedback on AI responses to guide future outputs  \\ \hline
Incremental refinement         & Iterative refinement on basic request based on AI responses \\ \hline
Multi-part problem solving     & Dividing the task into sub-problems addressed in sequence   \\ \hline
Single comprehensive prompt    & One prompt with all requirements, context, and constraints  \\ \hline
Step-by-step guidance          & Breaking problem down into sequential steps for AI to solve \\ \hline
Template-based approach        & Applying prompt templates adapted for task                  \\ \hline
\end{tabular}
}
\caption{Conversation strategy with GenAI tools}
\label{table:conv-strategy-desc}
\end{table}

\begin{figure}
    \centering
    \includegraphics[width=0.97\linewidth]{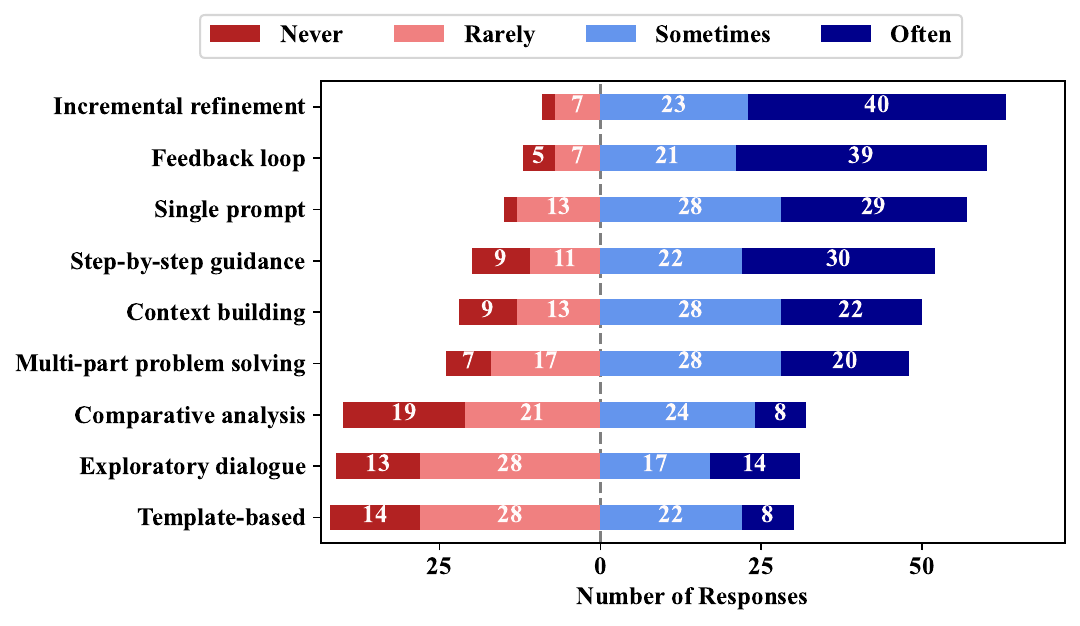}
    \caption{Conversation structure usage frequency}
    \label{fig:c1}
\end{figure}

\Cref{fig:c1} reveals that the most frequently used strategy was \strategy{Incremental refinement} (incrementally refining the same prompt) followed closely by \strategy{Feedback loop} (using potentially different prompts to achieve the desired result). These strategies reflect practitioners' preference for multi-turn conversations where they can guide and refine GenAI responses progressively. However, \strategy{Single comprehensive prompt} is the third most popular and does not reflect this iterative workflow. The least popular strategies were \strategy{Comparative analysis}, \strategy{Template-based approach}, and \strategy{Exploratory dialogue}. Despite their lower adoption rates, these strategies still showed meaningful usage patterns, with at least 40\% of practitioners employing them ``Sometimes'' or ``Often''.

\begin{figure}
    \centering
    \includegraphics[width=\linewidth]{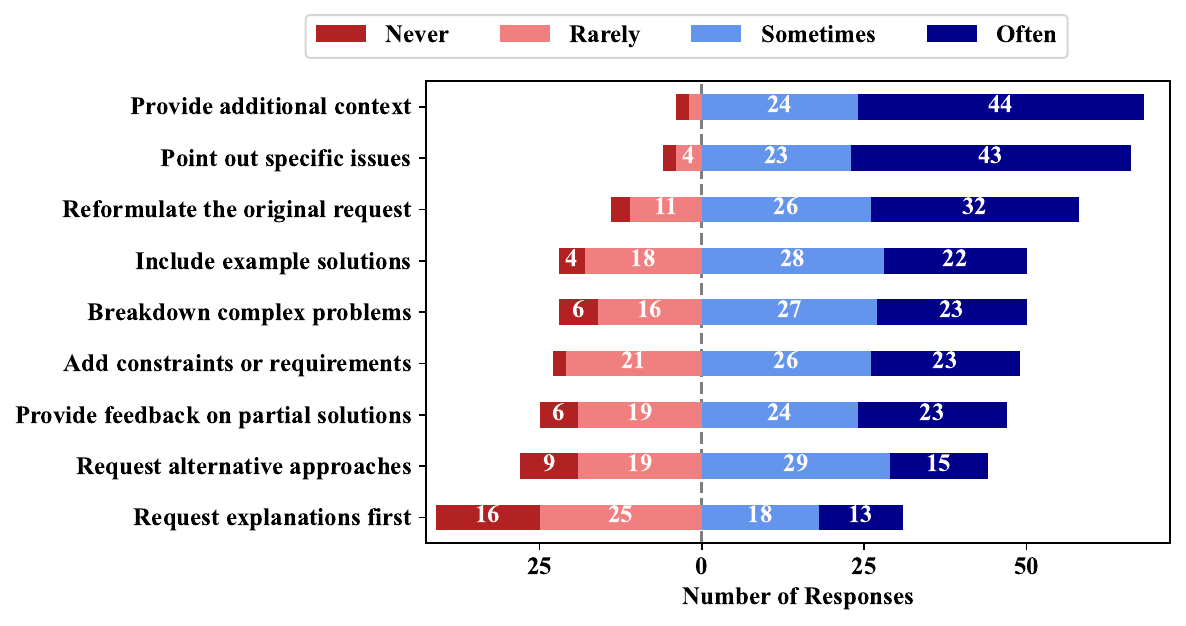}
    \caption{Error handling strategy usage frequency}
    \label{fig:c2}
\end{figure}

\subsubsection{Error Handling Strategies}
\label{sec:error-handling}

We asked participants to indicate which error handling strategies they use when they deem GenAI responses as inadequate. Participants were provided with nine strategies and could select none or any number of strategies. Practitioners converge on two complementary strategies (\Cref{fig:c2}): \textit{providing additional context} and \textit{pointing out specific issues}, with 79\% of frequent users employing both approaches. \Cref{fig:c2} shows that reformulating requests, adding constraints, including example solutions, breaking down complex problems, and providing feedback on partial solutions show more moderate adoption, while requesting alternatives or explanations remain far less popular. However, the 13 respondents who often use this least popular strategy are likely to use every other strategy: none chose ``Never'' in response to any option. There is a similar trend for the 15 respondents who frequently request alternatives: only three selected ``Never'' for requesting explanations and not for any other strategy. 

\finding{Practitioners who reported utilizing less popular error handling strategies typically also used all other provided strategies at least \textit{rarely}. This suggests that strategies are used as complements to others.}

We identified connections between conversation structuring and error handling approaches, focusing exclusively on users who responded ``Often'' to both types of strategies. 80\% of practitioners who frequently use \strategy{Incremental refinement} also provide additional context when errors occur. Similarly, 83\% of users who provide feedback on partial solutions frequently utilize \strategy{Feedback loop}. Those who use step-by-step guidance often break down complex problems when handling errors (50\%). These correlations show consistent practitioner strategies, which could inform future tool design: for example, a tool might activate additional context provision settings if a practitioner is aware of their preference for an iterative strategy, or adjust a prompting agent's writing style depending on the user.

\subsubsection{Exchange Count}
\label{sec:exchange_ct}

We asked respondents to report the average number of exchanges (back-and-forth interactions) needed to complete SE tasks with GenAI tools. Survey options included predefined ranges (1, 2-3, 4-6, 7-10, 10+, and Unsure). No participants reported consistently completing tasks in a single exchange, suggesting that current GenAI systems require iterative interaction regardless of user expertise or strategy choice. %
Self-reported GenAI skill levels show a positive correlation with exchange count, as seen in ~\Cref{table:skill-exchanges}. Maximally proficient users require substantially more exchanges than other skill levels, having by far the highest proportion of ``10+'' selections. Longer conversations also correlate with productivity gains: participants using 10+ exchanges were the only group to report exclusively positive productivity impacts, with 6/7 indicating ``significant'' improvements.

\finding{No practitioners reported consistently completing tasks with GenAI in a single exchange, while longer interactions appear associated with higher productivity and perceived proficiency.
}

\begin{table}[t]
\centering
\footnotesize
\resizebox{\columnwidth}{!}{
\begin{tabular}{|l|c|c|c|c|c|}
\hline
\textbf{GenAI Skill Level} & \textbf{2-3} & \textbf{4-6} & \textbf{7-10} & \textbf{10+} & \textbf{Unsure} \\
\hline
Somewhat Proficient & 3 & 7 & 0 & 0 & 4 \\
Proficient & 15 & 11 & 3 & 3 & 0 \\
Very Proficient & 5 & 9 & 1 & 1 & 2\\
Maximally Proficient & 1 & 3 & 0 & 3 & 0\\ \hline
Total & 24 & 30 & 4 & 7 & 6 \\
\hline
\end{tabular}
}
\caption{Exchange count distribution by self-reported skill}
\label{table:skill-exchanges}
\end{table}

\subsection{\ref{rq:3}: GenAI Reliability and Common Issues}

This section examines which SE subtasks practitioners perform using GenAI tools, their perceptions of these tools' reliability across SE scenarios, and how often they encounter various issues. Our results highlight where GenAI tools excel and struggle.%

\subsubsection{Task Engagement Patterns}

For each SE task, we analyzed which specific subtasks respondents reported using  GenAI for (\Cref{table:task_engagement}). The most common subtask for code generation was implementing new code components (52), followed by setting up structural code (34) and designing architectures (30). This pattern reflects a clear preference for using GenAI to assist with sub-tasks related to \textit{implementation} rather than \textit{design/analysis}. A similar implementation-focus emerges for some of the other SE tasks. For example, for testing, respondents showed a pronounced 6:1 preference for test creation/implementation (18) over test strategy/planning; and for documentation, a nearly 2:1 preference for code-level documentation (24) over project-level documentation (13). A few additional GenAI sub-tasks were not included in our predefined options, including reverse engineering, background research, and input generation for tests, each highlighted by one respondent.

\finding{For code generation, documentation, and testing, respondents generally used GenAI for implementation-focused (lower-level) sub-tasks rather than for design and analysis (higher-level) sub-tasks.}

\begin{table}[t]
\centering
\caption{Most frequently selected sub-tasks for each SE task. `I' = Implementation focus (lower abstraction level)\\ `D' = Design / Review focus (higher abstraction level)}
\label{table:task_engagement}
\resizebox{\columnwidth}{!}{
\begin{tabular}{lp{5cm}r}
\toprule
\textbf{SE Task} & \textbf{Sub-task} & \textbf{Count} \\
\midrule
Code Generation & Impl. new code components (I) & 52 \\
& Setting up structural code (I) & 34 \\
& Designing software architectures (D) & 30 \\
\midrule
Documentation & Code-level documentation (I) & 24 \\
& Project-level documentation (D) & 13 \\
\midrule
Testing & Test creation and implementation (I) & 18 \\
& Test strategy and planning (D) & 3 \\
\midrule
Debugging & Bug fixing (I) & 26 \\
& Issue diagnosis (D) & 22 \\
\midrule
Refactoring & Optimizing existing code (I) & 14 \\
& Standardizing and cleaning code (D) & 12 \\
\midrule
Review & Funct./perf. assessment (I) & 10 \\
& Code quality evaluation (D) & 7 \\
\bottomrule
\end{tabular}
}
\end{table}

\subsubsection{GenAI Perceived Reliability}

To assess practitioners' trust in GenAI tools, %
we analyzed reported reliability ratings for 12 scenarios spanning all six studied SE tasks. Participants rated each scenario on a 4-point scale from ``Very unreliable'' to ``Very reliable,'' with ``Unsure/Haven't tried'' responses excluded from our analysis. %

\begin{figure}[t]
    \centering
    \includegraphics[width=\linewidth]{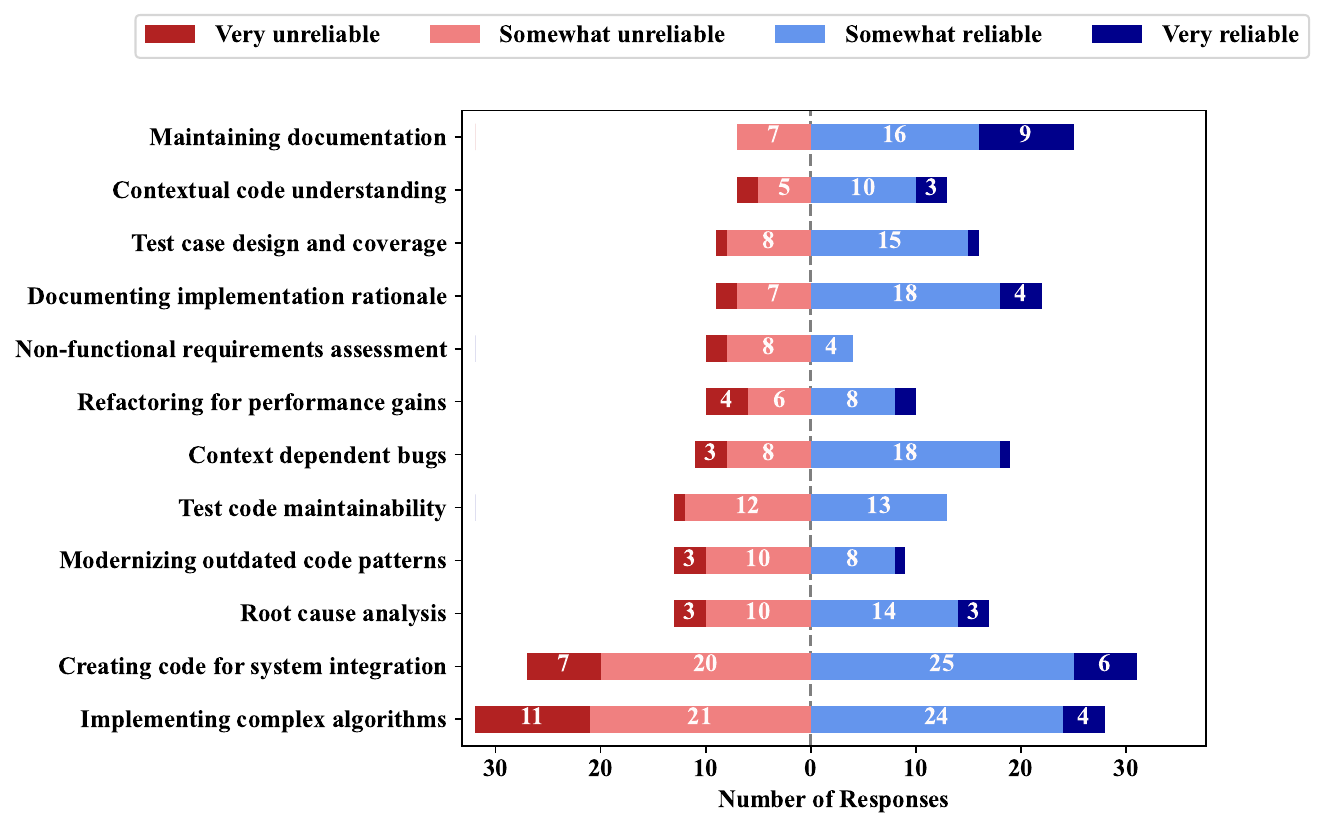}
    \caption{Reliability perceptions of GenAI across tasks}
    \label{fig:r2}
\end{figure}

\Cref{fig:r2} shows the perceived reliability of using GenAI for various tasks. Respondents reported that documentation tasks were the most reliable to complete with GenAI. In particular, maintaining documentation accuracy received the highest avg. rating (3.06, n=32), followed by documenting implementation rationale (2.77, n=31). Respondents showed moderate confidence in using GenAI for testing and debugging tasks, including test case design and coverage (2.64, n=25) and root cause analysis of existing errors (2.57, n=30). The tasks reported as least reliable by practitioners often involved complex analysis or reasoning. These included the assessment of non-functional requirements (2.14, n=14), the modernizing of outdated code patterns (2.32, n=22), and the implementation of complex algorithms (2.35, n=60). In particular, GenAI tools may struggle with non-functional requirements since many of these require a deep understanding of domain knowledge from a complex field, such as security or license compliance.

\finding{Respondents perceived GenAI as less reliable for software tasks requiring complex reasoning, analysis, and domain knowledge.}

\begin{table}[]
\resizebox{\columnwidth}{!}{
\begin{tabular}{|l|c|c|c|c|c|c|}
\hline
\textbf{Experience} & \textbf{Very-} & \textbf{Some-} & \textbf{Some+} & \textbf{Very+} & \textbf{Unsure} & \textbf{Score}\\ \hline
{Junior}          & 7.1\%  & 22.7\% & 44.8\% & 12.8\% & 12.7\% & 2.73      \\ \hline
{Intermed.}    & 12.2\% & 37.4\% & 36.0\% & 4.4\%  & 10.1\% & 2.36         \\ \hline
{Senior}          & 16.2\% & 35.3\% & 34.7\% & 2.6\%  & 11.2\% & 2.27      \\ \hline
{Veteran}         & 7.1\%  & 19.1\% & 47.6\% & 17.3\% & 8.9\% & 2.82       \\ \hline
\textbf{Proficiency} & \textbf{Very-} & \textbf{Some-} & \textbf{Some+} & \textbf{Very+} & \textbf{Unsure} & \textbf{Score}\\ \hline
{Some. Prof.}  & 29.5\% & 30.4\% & 27.4\% & 0\%     & 12.8\% & 1.98        \\ \hline
{Proficient}      & 6.2\%  & 33.1\% & 42.8\% & 11.3\% & 6.6\% & 2.63       \\ \hline
{Very Prof.} & 8.4\%  & 26.9\% & 37.7\% & 10.2\% & 16.7\% & 2.60           \\ \hline
{Max. Prof.} & 4.8\%  & 29.1\% & 46.7\% & 4.1\%  & 15.5\% & 2.59          \\ \hline
\end{tabular}
}
\caption{GenAI reliability results across GenAI proficiency and experience groups. ("Reliable" and "Unreliable" shortened to "+" and "-")}
\label{table:reliability}
\end{table}

GenAI reliability results across respondents' experience and self-reported GenAI proficiency groups can be seen in \Cref{table:reliability}. 
Participants were presented with subsets of the 12 reliability options from \Cref{fig:r2},  which varied based on the SE tasks they indicated using GenAI for. %
To account for this variation of presented options, 
we calculated proportions for each participant across all of their reliability responses, then calculated a reliability score as the mean of individual reliability ratings across each GenAI proficiency and experience category. Overall reliability scores were calculated by multiplying integer values for the Likert-scale options (``Very unreliable'' = 1, ``Very reliable'' = 4) by the calculated proportion of a given category, then dividing the \% of participants who gave a rating other than ``Unsure.'' Overall reliability proportions and scores are reported in \Cref{table:reliability}.  

Despite nearly equivalent totals for Proficient, Very Proficient, and Maximally Proficient users, Somewhat Proficient users have the lowest at 1.98, largely due to an outlying quantity of ``Very Unreliable'' selections. This is expected, since it is unlikely that respondents would assign themselves low proficiency if they were able to obtain reliable AI responses. The most proficient users were unlikely to classify GenAI as ``Very Unreliable'' or ``Very Reliable'' for any task, suggesting that they have at least occasionally witnessed GenAI fail in optimal scenarios and succeed in sub-optimal ones.

In contrast, our analysis by experience level shows no clear patterns. The most- and least-experienced respondents were more likely to find GenAI reliable than those in the middle. This may be correlated with Juniors' and Veterans' proportionally lower familiarity with prompting techniques, but that does not explain why Intermediates consider AI more reliable than Seniors. 

\finding{Senior practitioners (with 11-20 years of development experience) find GenAI least reliable on average, while veterans (with 21+ years) find it most reliable on average.}

\begin{figure}[t]
    \centering
    \includegraphics[width=\linewidth]{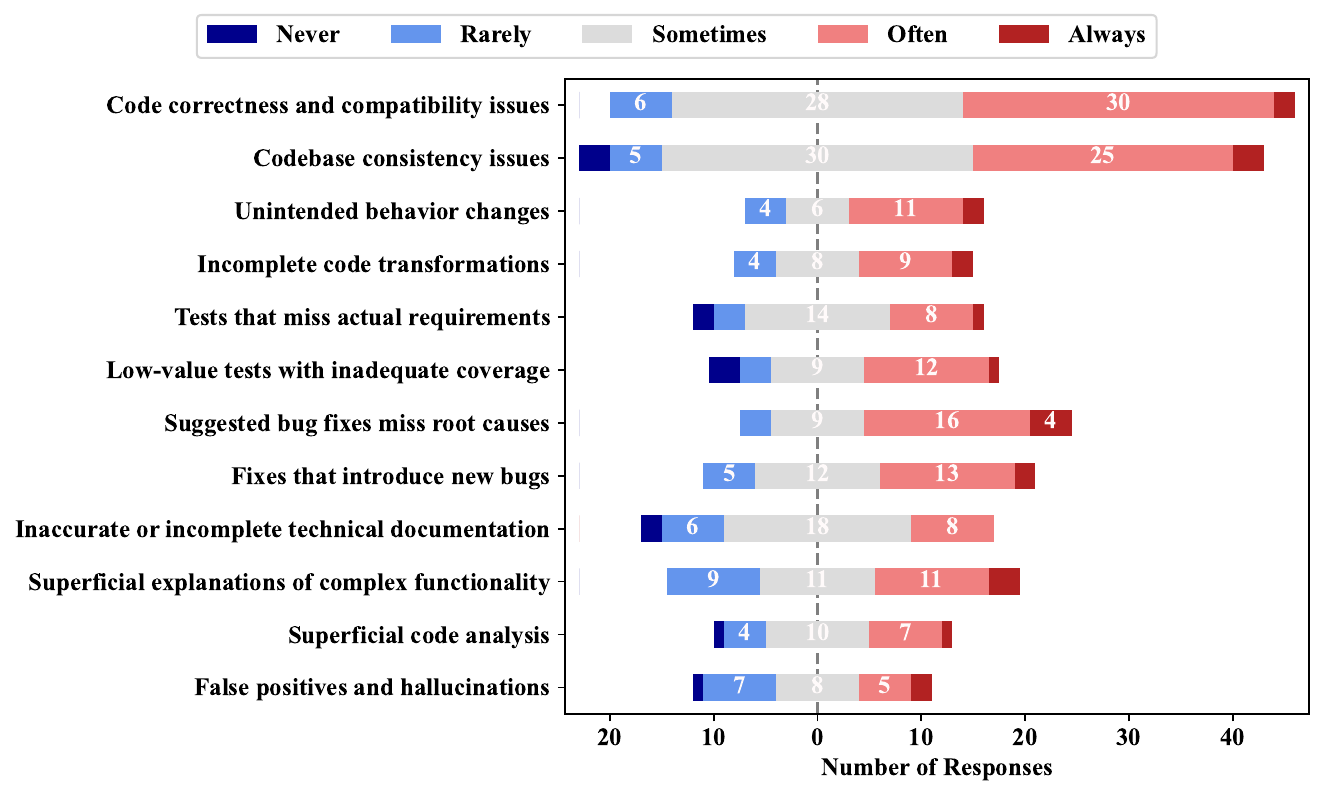}
    \caption{GenAI issue frequency distribution}
    \label{fig:r4}
\end{figure}

\subsubsection{Issue Frequency}
\label{sec:issue_frequency}

Our final reliability analysis examined how frequently practitioners encountered specific issues (given to them) when using GenAI tools across different SE contexts. Participants rated twelve task-specific issues on a five-point scale from ``Always'' (1) to ``Never'' (5), where lower mean scores indicate more frequent issue occurrence. Results are visualized in \Cref{fig:r4}.

Debugging emerges as the most problematic SE task, with suggested bug fixes missing the root cause having the highest issue frequency (2.34, n=32). This issue received four ``Always'' responses (the highest count for any absolute rating across all issues, despite only appearing for those who selected the  task), indicating that GenAI's diagnostic capabilities remain a critical hurdle. Refactoring and code generation follow in severity, with unintended changes to behavior (2.52, n=23) and correctness/compatibility issues in code (2.58, n=66) representing significant challenges. Documentation and code review showed the lowest reported issue frequencies, with practitioners noticing inaccurate or incomplete technical documentation (3.06, n=34) and false positives and hallucinations (3.00, n=23) least often. This aligns with our previous findings about the lower complexity of these tasks.

\finding{Respondents reported experiencing the most issues when using GenAI for debugging, particularly when it fails to provide root causes. Code generation and refactoring were also noted pain points.}

\begin{table}[t]
\centering
\caption{GenAI issue frequency ranking across SE tasks
} 
\small
\label{tab:issue_frequency}
\begin{tabular}{cp{5.5cm}cc}
\toprule
\textbf{Rank} & \textbf{Issue} & \textbf{Mean} & \textbf{n} \\
\midrule
1 & Suggested bug fixes miss root causes & 2.34 & 32 \\
2 & Unintended behavior changes & 2.52 & 23 \\
3 & Code correctness and compatibility issues & 2.58 & 66 \\
4 & Incomplete code transformations & 2.61 & 23 \\
5 & Fixes that introduce new bugs & 2.63 & 32 \\
6 & Codebase consistency issues & 2.70 & 66 \\
7 & Superficial explanations of functionality & 2.76 & 34 \\
8 & Low-value tests with inadequate coverage & 2.82 & 28 \\
9 & Superficial code analysis & 2.87 & 23 \\
10 & Tests that miss actual requirements & 2.89 & 28 \\
11 & False positives and hallucinations & 3.00 & 23 \\
12 & Inaccurate or incomplete documentation & 3.06 & 34 \\
\bottomrule
\end{tabular}
\end{table}

\section{Discussion}
\label{sec:disc}

\subsection{Developing Proficiency with Generative AI}

Our results show a positive relationship between a practitioner's perceived GenAI proficiency and the number of tasks they complete with GenAI (\Cref{table:skill-vs-tasks}), suggesting that task diversity and GenAI proficiency may mutually reinforce each other. This could potentially reflect how confidence encourages practitioners to explore GenAI usage in new contexts, or could also demonstrate that using GenAI on a wide range of tasks increases proficiency.

Users' prompt construction strategies vary significantly among tasks. Respondents showed a high consensus in the information they reported providing to models when completing debugging tasks, with 94\% including full error messages/stack traces. By contrast, there was little agreement on the information included for documentation tasks, despite those being the second most common SE tasks completed with GenAI assistance: only 71\% (24/34) of respondents reported providing the model with the code snippet and a functionality explanation. Code Generation has the next lowest with 76\% of respondents including example inputs and expected outputs. All tasks show {some} convergence of information inclusion strategies, and debugging and testing (which have the highest agreement) explicitly involve work on existing code, making that necessary to share in most instances. However, Documentation also involves work on existing code in many cases, suggesting that consensus differences have no clear pattern.

One counterintuitive finding is that maximally proficient users require substantially more exchanges to complete tasks than less skilled users (\Cref{table:skill-exchanges}), which seems to indicate that higher skill correlates with inefficiency. Another interpretation is that the strategies employed by these high-skill users are not inefficient, but naturally result in more prompts by enabling users to tackle more ambitious problems. Notably, longer conversations correlate with productivity gains: participants using 10+ exchanges were the only group to report exclusively positive productivity impacts, with 6/7 indicating ``significant'' improvements. This suggests that sustained engagement with GenAI tools enables more meaningful outcomes, rather than expecting immediate solutions. %

\subsection{GenAI Limitations \& Practical Implications}

Debugging does not have the lowest overall reliability perception, but it is prominent as the task with the highest issue frequency (\#1 and \#5 on \Cref{tab:issue_frequency}), while other functional coding problems (refactoring and generation) fill the top 50\%. This suggests consistent limitations in GenAI's ability to reason about code behavior and the effects of modifications, compared to natural language tasks such as documentation and review. %

While these results provide some direction for the critical problems GenAI engineers should focus on solving, others point toward alternative ways to improve reliability and user experience. \Cref{table:skill-exchanges} shows how the number of GenAI exchanges increases with a user's skill, and users tend to prefer iterative conversations (\Cref{sec:error-handling}). Accordingly, high-profile tools usually facilitate iteration in some respect, most obviously by allowing users to send multiple prompts throughout a conversation. However, there are other ways to facilitate iteration, such as allowing users to edit earlier prompts to create conversation branches. %

Our findings indicate that daily use makes productivity improvements significantly more likely (\Cref{sec:productivity}), and also correlates with GenAI use on a broader range of tasks. This suggests that the best way for individual practitioners to realize the benefits of these new tools is to carefully consider which tasks in their workflow could benefit from utilizing GenAI, then purposefully learn about and apply GenAI there. Gradual adjustment may be necessary to ensure consistent performance and build an understanding of how best to navigate AI limitations.

\subsection{High Technique Awareness, Low Adoption}
Our survey population had an overwhelming familiarity with various prompting techniques, with their awareness of the \strategy{Meta Language Creation} technique being the lowest (74\%).  However, despite an overall high awareness, there was a lackluster adoption.  For example, even though 92\% (66/72) of respondents indicated familiarity with the \strategy{Output Style} technique, only 82\% of those (54) actually used it at all, and of those only 54\% (29) used it regularly. The adoption percentages are worse for the remaining techniques, and regular usages are well below 50\%.  This is particularly interesting given that awareness of research outcomes is typically one of the biggest hurdles for tool and technique adoption, meaning that the issue must lie elsewhere. 
It may be that either practitioners do not find these techniques helpful or that they are unaware of how to effectively use them. 
The results in \Cref{table:skill_vs_techniques} suggest that decades of work experience may assist veteran practitioners in finding use cases for prompting techniques, since this was the group most likely to adopt these techniques.  However, flexibility may also be a crucial component to effective use, as potentially demonstrated by the intermediate practitioners who had the highest frequency of regular technique usage. 
Accordingly, future research can further explore the pain points that practitioners encounter when applying these techniques to their workflows.  Additionally, a comprehensive taxonomy and/or usage guide could potentially assist practitioners in more effectively using these prompting techniques or understanding how to apply them to their particular use cases.

\subsection{Future Work}

While existing work provided a helpful starting point for the prompting techniques included in the survey, it was not comprehensive enough to account for all interactions, necessitating our definitions of \strategy{Output Style} and \strategy{Condition Check}. Despite these additions, our qualitative analysis revealed strategies not adequately represented in the survey. 
One user instructed the GenAI to avoid outputting explanations of its work, while another deliberately excluded examples of input and output from prompts, suggesting that some users develop preferences for concise interactions or that extra context can reduce model output quality in certain situations. 

To rigorously study the effectiveness, variation, and progression of GenAI interactions, a comprehensive taxonomy of all possible interaction types must be created. This would enable systematic categorization and quantification of strategies, allowing controlled comparisons of effectiveness across tasks and user populations.

Given iteration's significant correlation with skill and productivity, investigating further iterative interaction methods could be valuable. For example, allowing users to edit model parameters (temperature, output length, \etc) mid-conversation could open new solution paths and grant greater control of GenAI tools. %

\section{Threats to Validity}
\label{sec:threats}

\textbf{Internal Validity.}
To mitigate possible coding biases, we used a rigorous iterative open-coding methodology to analyze open-ended survey responses. We also followed the best practices when formulating our survey questions in order to avoid confusing and/or biasing language, and conducted a small pilot study with graduate students to test the clarity and length of our questionnaire.  We recruited participants from a variety of sources (our professional networks, university organizations, \etc) to ensure we received a broad range of perspectives and experiences, but we are aware of the threat of self-selection bias. 

\textbf{External Validity.}
Study conclusions apply only to participants who completed our survey and cannot be generalized to larger practitioner populations. However, given the themes and trends that emerged, we believe that many practitioners will share similar experiences to those reported.  Ultimately, our goal was never to claim generalizability, but to explore the emerging landscape of how practitioners are interacting with GenAI tools for SE tasks.

\textbf{Construct Validity.}
This work relies on self-reported practitioner perceptions, which can be subjective and influenced by recall and social desirability biases. We sought to mitigate this threat by corroborating and anchoring subjective assessments such as proficiency with more concrete metrics captured within the survey, including the breadth of AI-assisted tasks participants perform and their frequency of tool use, and avoiding making definite claims about practices. %

\section{Related Work}
\label{sec:back}

Generative AI models are systems capable of creating new content based on their training data. Although the definitions of `prompting' vary, they generally refer to providing GenAI with input, usually text instructions and / or an image, to guide the model toward producing the intended result. Upon release, these tools were quickly applied to SE, with an early study by GitHub on their Copilot integration \cite{github2022copilot} showing significant gains in coding speed, job satisfaction, and perceived productivity. A more recent mixed-methods study from Butler et al. \cite{butler2024deardiaryrandomizedcontrolled} within a large software company showed an increase in positive views toward AI coding tools, but without a corresponding boost to developer trust, which is notable because trust factors play an important role in developer intentions \cite{choudhuri2024guideschoicesmodelingdevelopers} \cite{choudhuri2025needsattentionprioritizingdrivers}. The utility of AI tools goes beyond coding tasks, as APIs or locally run models have led to a new design paradigm of integrated, interacting commands with more traditional structured code \cite{liang2025promptsprogramstoounderstanding, tafreshipour2025promptingwildempiricalstudy}. Multiple reviews of LLM \cite{hou2024largelanguagemodelssoftware, fan2023largelanguagemodelssoftware} and LLM-based agents \cite{liu2024largelanguagemodelbasedagents} for SE show extensive possibilities to further improve SE workflows.

This has created the field of prompt engineering, which serves to determine the most effective ways of prompting a model in order to produce high-quality output which suits the user's needs. In its early stages, this resulted in explorations of individual strategies, leading to catalogs of prompt patterns \cite{white2023promptpatterncatalogenhance} applicable to a broad range of interaction types \cite{treude2025developers}. As more patterns were refined, improved and discovered, compilations of techniques were formed through extensive reviews of the academic literature \cite{schulhoff2025promptreportsystematicsurvey, sahoo2025systematicsurveypromptengineering}, though empirical analysis has found no significant differences in code quality across basic prompt patterns \cite{Della2025prompt}. Others attempted to apply prompt engineering for specific applications such as security \cite{tony2025promptingtechniquessecurecode}, design more complex conversational workflows \cite{robino2025conversationroutinespromptengineering}, and even use the generative AI models themselves for prompt engineering tasks \cite{zhou2023largelanguagemodelshumanlevel}. With new advances such as chain-of-thought outputs \cite{treude2025interacting}, human-AI interaction is evolving beyond just prompting.

Although these frameworks and pattern collections have provided an important foundation for understanding prompting capabilities, they reveal little about how software developers use AI tools in practice. The DevGPT dataset \cite{Xiao_2024} and various analyses of it have sought to close this gap, but often focus on a broader analysis of collaborative contexts \cite{hao2024empiricalstudydevelopersshared}, specific tasks such as issue resolution \cite{ehsani2025detectingpromptknowledgegaps}, or high-level usage patterns such as autocomplete / coding acceleration \cite{10548213} rather than an examination encompassing diverse types of SE tasks. Li et al. conducted a comprehensive study on the individual and organizational motives of AI adoption and use among software developers \cite{li2024ai}, while Simaremare et al. highlighted the key benefits and challenges of doing so \cite{Simaremare2024state}. Other approaches involve analysis of discourse from social media sites such as X/Twitter \cite{BASHA2025107804} or user reviews about AI coding extensions \cite{lyu2025myproductivityboosted}, while surveys of AI adoption among developers \cite{10.1145/3652154, lambiase2025exploringindividualfactorsadoption, zakharov2025aisoftwareengineeringperceived} in Sweden \cite{Lakis_Rifai_2025}, Slovenia \cite{robas2025slovenia}, and Indonesia \cite{anditama2025indonesia} reveal factors influencing use and perceptions among certain populations. 

While many distinct techniques have been researched and compiled, there has been no investigation comparing practitioners' preferred prompting strategies and interaction methods across multiple categories of software engineering tasks, leaving information about how these tools are used in practice relatively scarce. Our study addresses this gap with a comprehensive analysis of prompting and conversation practices in six distinct SE domains.

\section{Conclusion}
\label{sec:conc}

We conducted a comprehensive survey of 72 software practitioners to investigate how they integrate GenAI tools into software engineering workflows. Our findings reveal primary adoption of GenAI for code generation, while use for related tasks such as documentation and debugging strongly correlates with a practitioner's AI proficiency and usage frequency. Users tend to prefer iterative conversations to achieve their goals, with consistent single-shot prompting remaining an impossibility. Significant reliability challenges remain for complex coding tasks, particularly debugging.

\bibliographystyle{ACM-Reference-Format}
\bibliography{references}

\end{document}